\documentclass{PoS}
\usepackage{graphicx}
\newcommand{\msol}{M$_\odot$}

\newcommand{\hi}{\mbox{H{\sc i}}}
\newcommand{\kms}{km s$^{-1}$}

\title{Deep wide field \hi\ imaging of M31}

\ShortTitle{Deep wide field HI imaging of M31}

\author{\speaker{Laurent Chemin}\\
    G\'EPI, Observatoire de Paris, section Meudon, CNRS \& Universit\'e
    Paris 7 UMR 8111, 5 Place Jules Janssen, 92195 Meudon, France\\
    E-mail: \email{laurent.chemin@obspm.fr}}
\author{Claude Carignan\\
    Laboratoire d'Astrophysique Exp\'erimentale (LAE),
    Observatoire du mont M\'egantic, and D\'epartement de physique,
    Universit\'e de Montr\'eal, C.P. 6128, Succ. Centre-Ville,
    Montr\'eal, QC, Canada H3C 3J7\\
    Observatoire d'Astrophysique de l'Universit\'e de Ouagadougou (UFR/SEA),
    03 BP 7021 Ouagadougou 03, Burkina Faso}
\author{Tyler Foster\\
    Department of Physics \& Astronomy, Brandon University,
    Brandon, MB, Canada R7A 6A9}
 
\abstract{We report on preliminary results from a
 new deep 21-cm survey of the Andromeda galaxy, based on observations performed with the Synthesis Telescope and 
the 26-m antenna at DRAO. The  \hi\ distribution and kinematics of the disc are analyzed 
and basic dynamical properties are derived.
New \hi\ structures are discovered, 
like thin \hi\ spur-like structures and an external arm in the disc outskirts. 
The \hi\ spurs are related to perturbed stellar clumps outside the main disc of M31. 
The external arm lies on the far, receding side of the galaxy and has no obvious counterpart in the opposite side. 
These \hi\ perturbations probably result from tidal interactions with companions. 
It is found a dynamical mass of  $(4.7 \pm 0.5) \times 10^{11}$ \msol\ enclosed within a radius $R = 38$ kpc and 
a total mass of $\sim 1 \times 10^{12}$ \msol\ inside the virial radius.}

\FullConference{Panoramic Radio Astronomy: Wide-field 1-2 GHz research on galaxy evolution - PRA2009\\
		 June 02 - 05 2009\\
		 Groningen, the Netherlands}

\begin{document}

\section{Introduction}
Understanding the formation and evolution of galaxies like the Milky Way is one of the major goal of astrophysics.  
The Andromeda galaxy (M31) is very well suited to put constrains on the physical processes 
that control the evolution of spiral galaxies because of its proximity. 
The important stellar features related to the evolution of M31 are the faint, extended and perturbed  structures 
seen in the disc outskirts, in addition to many of the dwarf companions that have been detected in its close neighbourhood (\cite{iba07}). 
They are undoubtedly the imprints of the hierarchical growth of the stellar disc and halo of M31,   
similar to those seen in numerical models of dark matter and galaxy evolution in the framework of 
the Cold Dark Matter paradigm (e.g. \cite{spr05}). 

In this work we are interested in providing a detailed view of the neutral hydrogen disc of M31 from 
recent, deep, wide-field and high angular \hi\ imaging. Our direct objectives 
are (i) to study the most extended \hi\ distribution of M31, (ii) to derive an accurate \hi\ rotation curve for it, and 
(iii) to derive its basic dynamical parameters in order to put further constraints on the history of its mass assembly.  
 A few preliminary results are described herefter. A complete analysis of the data are presented in Chemin et al. (2009, \cite{che09}). 

\section{Observations}
The \hi\ observations were performed with the Synthesis Telescope and the 26-m dish at the Dominion Radio
Astrophysical Observatory (DRAO) between September and December 2005. Five fields were observed in the direction of M31 
for a total exposure time of 144 hours per field. The spectral resolution is 5.3 \kms\ and the angular size 
of the synthesized beam is $\sim 60" \times 90"$. It samples a linear scale of $\sim 230 \rm pc \times 340$ pc 
at the M31 distance (785 kpc, \cite{mcc05}).

\section{Results}

\begin{figure}[!t]
\centering
\includegraphics[width=6.9cm]{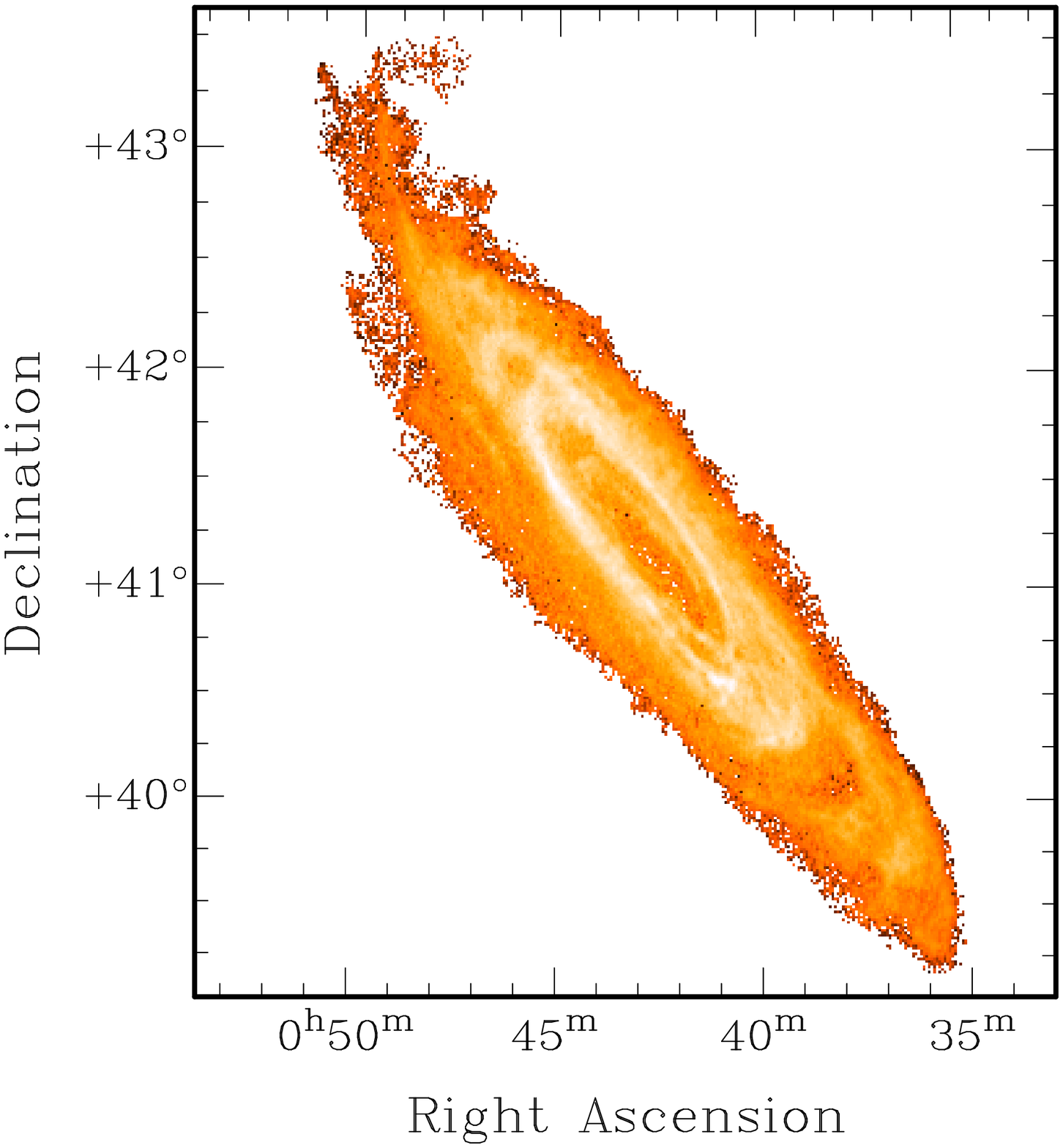}\includegraphics[height=7.53cm]{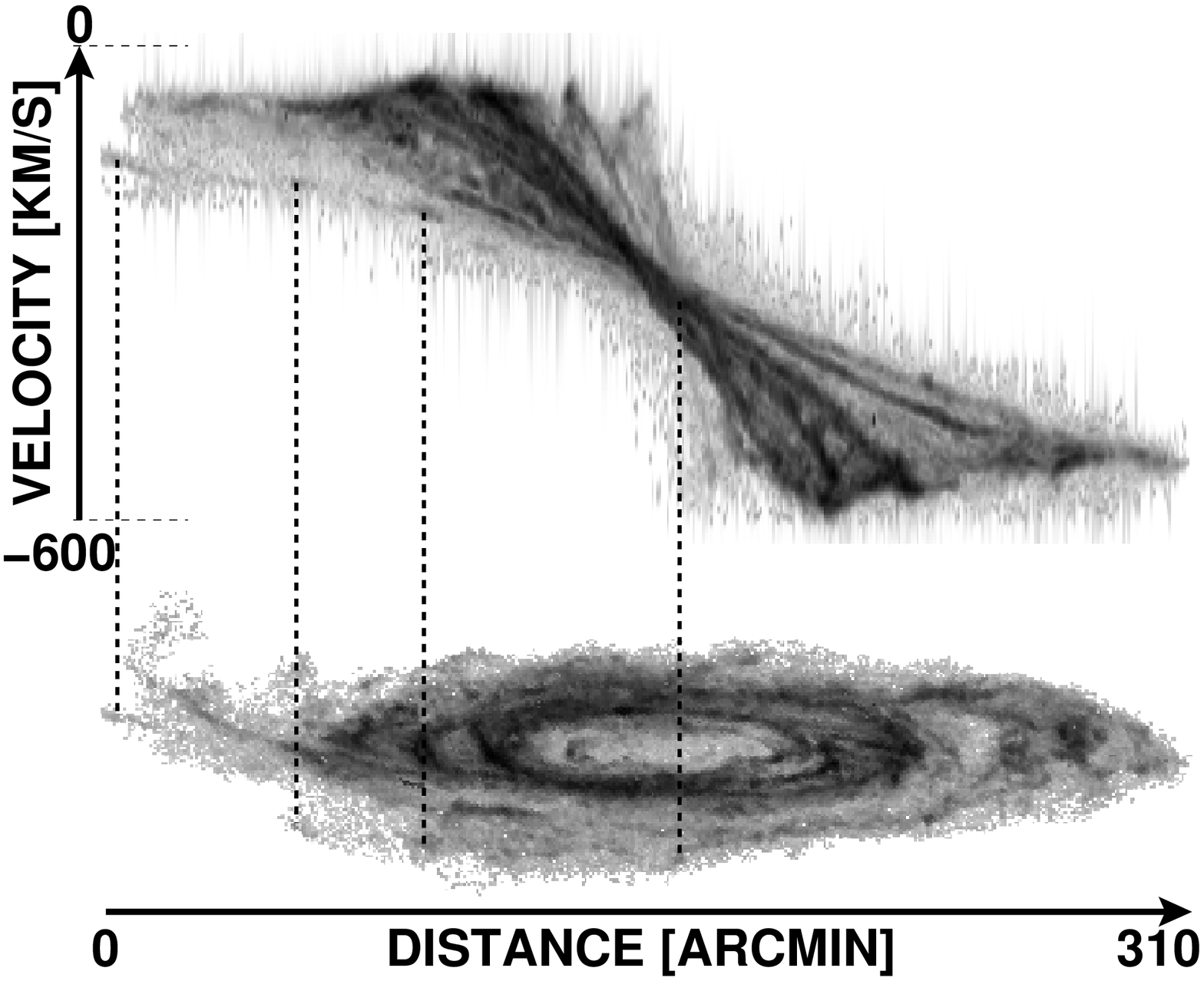}
\caption{\textbf{Left-hand panel:} \hi\ integrated emssion of Messier 31. \textbf{Right-hand panel:} 3D view of the \hi\ datacube of M31.  
The top panel is the position-velocity diagram of the full datacube projected 
onto the photometric major axis. The bottom panel is the same map as in the right-hand panel but displayed with the major 
axis parallel to the horizontal axis.  Dashed lines show the location of the newly discovered external arm (see text for details).}
\label{im1}
\end{figure}

The integrated emission is displayed in Figure~\ref{im1} (left-hand panel).
The high resolution \hi\ map shows a disc with very little gas in its central regions, 
as ususally observed in early-type discs. Faint spiral-like or ring-like structures are observed 
at $R \sim 2.5$ kpc  and $R \sim 5$ kpc.  They coincide with dusty ring-like structures observed in NIR images 
from SPITZER-IRAC data (\cite{bar06}) as well as with molecular gas ring-like structures (\cite{nie06}). 
Other brighter spiral- or ring-like structures are observed between $R \sim 9$ kpc and 
$R \sim 18$ kpc. Part of these \hi\ structures have already been presented in previous studies of M31 (\cite{bri84, unw83}).

New faint structures that were not seen in old \hi\ images are discovered. 
First the two disc extremities  exhibit thin spur-like extensions, particularly towards the North-East.  
Their kinematics are in good continuity to the adjacent inner disc. 
Velocity gradients are detected in these spurs ($\sim 20$ \kms\ along $\sim 7-10$ kpc). 
These spurs appear tightly linked to stellar clumps (the ``G1" clump and the NE extension, as identified in \cite{iba07}).

Then an external spiral arm is discovered on the edge of the receding half of the disc. 
It is outlined with dashed lines in Figure~\ref{im1} (right-hand panel). 
The \hi\ mass of this new arm is $\sim 10^8$ \msol. Its apparent length is $\sim 32$ kpc.
 It is connected to another more extended, brighter spiral arm. 
 The arm is clumpy towards its northeastern end. 
 Part of the thin NE \hi\ spurs is a kinematical extension of that external arm, as seen in the postion-velocity plot of  
 Fig.~\ref{im1}.  
 It is striking that the external arm has no evident morphological and kinematical counterpart in the approaching half of 
 the disc with respect to the galactic centre. 
 Moreover, it is obvious that its kinematics is very peculiar compared to the disc velocities.
 As a consequence it is very likely that this spiral-like structure has been generated by external effects to M31, 
 either by tidal effects after the passage of a galaxy satellite or even by gas accretion from the intergalactic medium or from a 
 former gas rich companion (NGC 205?). It is also remarkable how this external structure  
 has  a similar   mass as the Magellanic Stream in the halo of the Milky Way (\cite{put98}).
   Numerical simulations are necessary to investigate the different possibilities 
 and find the origin of the faint new perturbations. 
 Notice that we are confident about their detection because they are also observed in the recent 
 \hi\ map of M31 by Braun et al. (2009, \cite{bra09}). 
  
\section{Dynamical analysis}
Our previous dynamical analysis of M31 allowed us to derive a total mass of 3.4 $\times 10^{11}$ \msol\ for 
$R < 35$ kpc from single dish data obtained at  Effelsberg and GBT (\cite{car06}).
A new, more extended rotation curve is derived from a tilted-ring analysis of the velocity field of M31. A mass distribution model 
is fitted to the rotation curve. As provisional results, 
it is derived a dynamical mass  of $\rm M_{\rm Dyn} = (4.7 \pm 0.5) \times 10^{11}$ \msol\ 
 inside $R = 38$ kpc and a  dark-to-baryonic mass ratio of
 $\rm M_{\rm Dark}/M_{\rm Baryon} \sim 4.0 $ (79\% of dark matter, 21\% of luminous baryons). 
  Here $\rm M_{\rm Baryon}$ represents the sum of the black hole, gaseous and stellar masses, 
$\rm M_{\rm Dark}$ the dark matter mass.  
The total mass of M31 extrapolated to the virial radius  ($R \sim 160$ kpc) 
is $M_{\rm Vir} \sim 1.0 \times 10^{12}$ \msol. 
All these measurements are in excellent agreement with  results found from other dynamical tracers (\cite{eva00}, \cite{iba04}) and 
from another recent \hi\ survey (\cite{cor10}).
A very good concensus seems to have been obtained for the enclosed mass inside the inner 38 kpc of M31, as well as for its 
total mass. We refer to Chemin et al. (2009, \cite{che09}) for a more detailed discussion 
of these results.

\section*{Acknowledgements}
 We are very grateful to the staff of the Dominion Radio Astronomy Observatory at Penticton  for their support in
obtaining the observations, and especially T. Landecker for encouraging us to pursue this project. 
We thank the scientific and local committees at ASTRON and Groningen for having organized 
the PRA 2009 conference and for all stimulating discussions during the conference.


\begin{thebibliography}{99}
\bibitem{bar06} Barmby P., Ashby M. L. N., Bianchi L., Engelbracht C. W., Gehrz R. D., Gordon K. D., Hinz J. L., Huchra J. P., et al. 2006,
ApJ, 650, L45
\bibitem{bra09} Braun R., Thilker D. A., Walterbos R. A. M., 
\& Corbelli E.  2009, ApJ, 695, 937
\bibitem{bri84} Brinks E., \& Shane W. W. 1984, A\&ASS, 55, 179
\bibitem{car06} Carignan C., Chemin L., Huchtmeier W. K., \& Lockman F. J. 2006, ApJ, 641, L109
\bibitem{che09} Chemin L., Carignan C., Foster T., 2009, ApJ, 705, 1395
\bibitem{cor10} Corbelli E., et al., 2010, A\&A, in press, arXiv0912.4133
\bibitem{eva00} Evans N. W., Wilkinson M. I., Guhathakurta P., Grebel E. K., \& Vogt S. S. 2000, ApJ, 540, L9
\bibitem{iba04} Ibata R., Chapman S., Ferguson A. M. N., Irwin M., \& Lewis G.,  2004, MNRAS, 351, 117
\bibitem{iba07} Ibata R., Martin N. F., Irwin M., Chapman S., Ferguson A. M. N., Lewis G.,  \& McConnachie A. W. 2007, ApJ, 671, 1591
\bibitem{mcc05} McConnachie A. W., Irwin M. J., Ferguson R. A., Ibata R. A., Lewis G. F., \& Tanvir N. 2005, MNRAS, 356, 979
\bibitem{nie06} Nieten C., Neininger N., Gu\'elin M.,  Ungerechts H., Lucas R., Berkhuijsen E. M.,
Beck R., \& Wielebinski R., 2006, A\&A, 453, 459
\bibitem{put98} Putman  M. E., et al. 1998, Nature, 394, 752
\bibitem{spr05} Springel V., White S. D. M., Jenkins A., Frenk C. S., Yoshida N., Gao L., Navarro J. F., Thacker R., et al., 2005, Nature, 435, 629
\bibitem{unw83} Unwin S. C. 1983, MNRAS, 205, 787
 
\end{thebibliography}
\end{document}